\documentclass[10pt,twocolumn,letterpaper]{article}

\usepackage{iccv}
\usepackage{times}
\usepackage{epsfig}
\usepackage{graphicx}
\usepackage{amsmath}
\usepackage{amssymb}
\usepackage{pifont}
\usepackage[accsupp]{axessibility} 

\newcommand{\xmark}{\ding{55}}%
\usepackage{booktabs}
\usepackage{caption}

\usepackage{hyperref}
\hypersetup{breaklinks=true,colorlinks=true}

\iccvfinalcopy 

\def\iccvPaperID{3951} 

\ificcvfinal\fi

\begin{document}

\title{CodeNeRF: Disentangled Neural Radiance Fields for Object Categories}

\author{
Wonbong Jang \qquad 
Lourdes Agapito \\
Department of Computer Science\\
University College London\\
{\tt\small
    \{ucabwja,l.agapito\}@ucl.ac.uk
}

}


\twocolumn[{%
  \renewcommand\twocolumn[1][]{#1}%
\maketitle
\thispagestyle{empty}
\begin{center}
  \newcommand{\teaserwidth}{\textwidth}
  \centerline{
    \includegraphics[width=\teaserwidth]{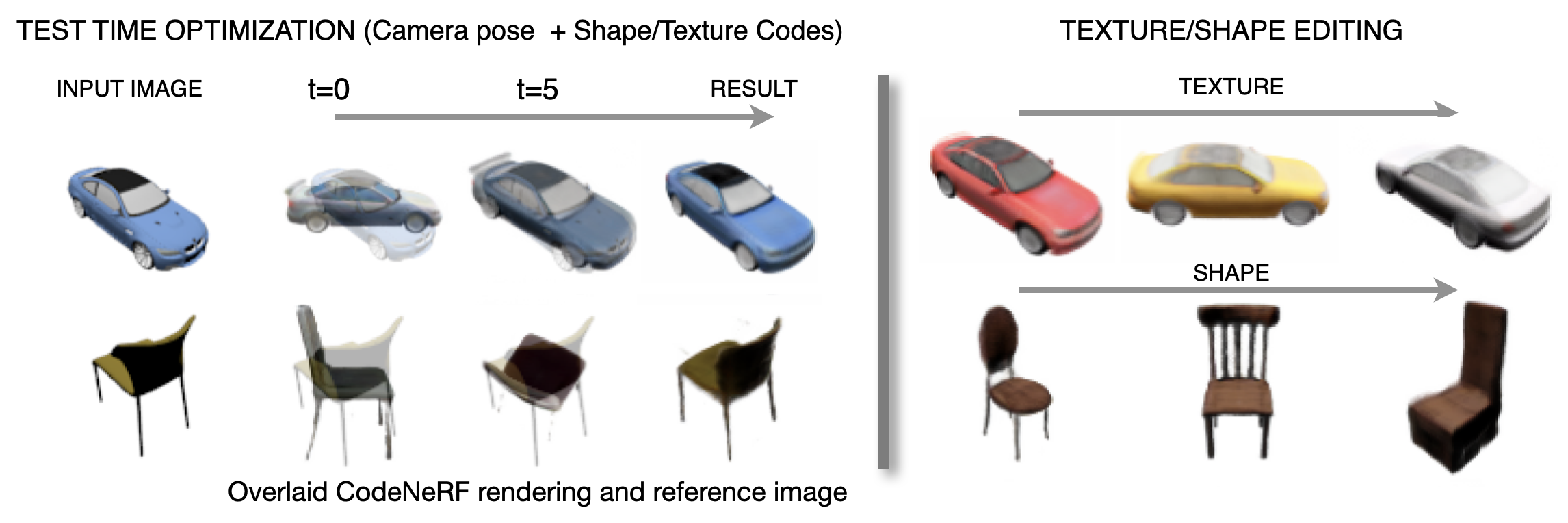}
    }
    \captionof{figure}{CodeNeRF represents variations of shape and texture across an object class.  \textbf{Left}: At test time, given a single input image, the trained model can be used to jointly optimize camera viewpoint and shape/texture latent codes. \textbf{Right}: Object shapes, textures and viewpoints can later be edited simply by varying the latent codes, offering full control over synthesis.}
  \label{fig:teaser}
 \end{center}%
}]

\begin{abstract}

CodeNeRF is an implicit 3D neural representation that learns the variation of object shapes and textures across a category and can be trained, from a set of posed images, to synthesize novel views of unseen objects. Unlike the original NeRF, which is scene specific, CodeNeRF learns to disentangle shape and texture by learning separate embeddings. At test time, given a single unposed image of an unseen object, CodeNeRF jointly estimates camera viewpoint, and shape and appearance codes via optimization. Unseen objects can be reconstructed from a single image, and then rendered from new viewpoints or their shape and texture edited by varying the latent codes. We conduct experiments on the SRN benchmark, which show that CodeNeRF generalises well to unseen objects and achieves on-par performance with methods that require known camera pose at test time. Our results on real-world images demonstrate that CodeNeRF can bridge the sim-to-real gap. Project page: \url{https://github.com/wayne1123/code-nerf}
\end{abstract}

\section{Introduction}

Synthesizing novel views of unseen objects given a sparse set of input views or even a single image is a long-standing problem in the fields of computer vision and graphics. 
Synthesis methods require both an accurate representation of the 3D geometry and appearance of objects, and the ability to offer control over changes in viewpoint, shape or texture to render different objects of the same category. 

Traditionally, discrete scene representations have been used, such as  meshes or voxel grids that store geometry and appearance information either explicitly or via learnt neural features~\cite{sitzmann2019deepvoxels}. However, their discrete nature limits their representation power and  resolution. With the recent introduction of scene representation networks (SRN), Sitzmann \etal~\cite{sitzmann2020scene} propose to learn a continuous function that map 3D  locations to features of scene properties. Crucially, SRN does not require  3D supervision and can be trained end-to-end with a differentiable ray marcher that renders the feature-based representation into a set of 2D images. While SRN allows generalization to unseen objects, accurate camera pose estimates are required at inference time, and shape and appearance are represented in an entangled manner. 

Mildenhall \etal~\cite{mildenhall2020nerf} extended the neural representation to store volume density and view-dependent radiance values, enabling highly photo-realistic new view synthesis of complex real-world scenes that capture viewpoint dependent effects. New view synthesis is done by querying the network at each specific spatial location and viewing direction, followed by classic volume rendering to output image pixel intensities. 
While the quality of synthesized images is impressive, it requires a large number of  images, and must be optimized for each new scene independently. 
\begin{figure}[t]
\centering
\includegraphics[width=.45\textwidth]{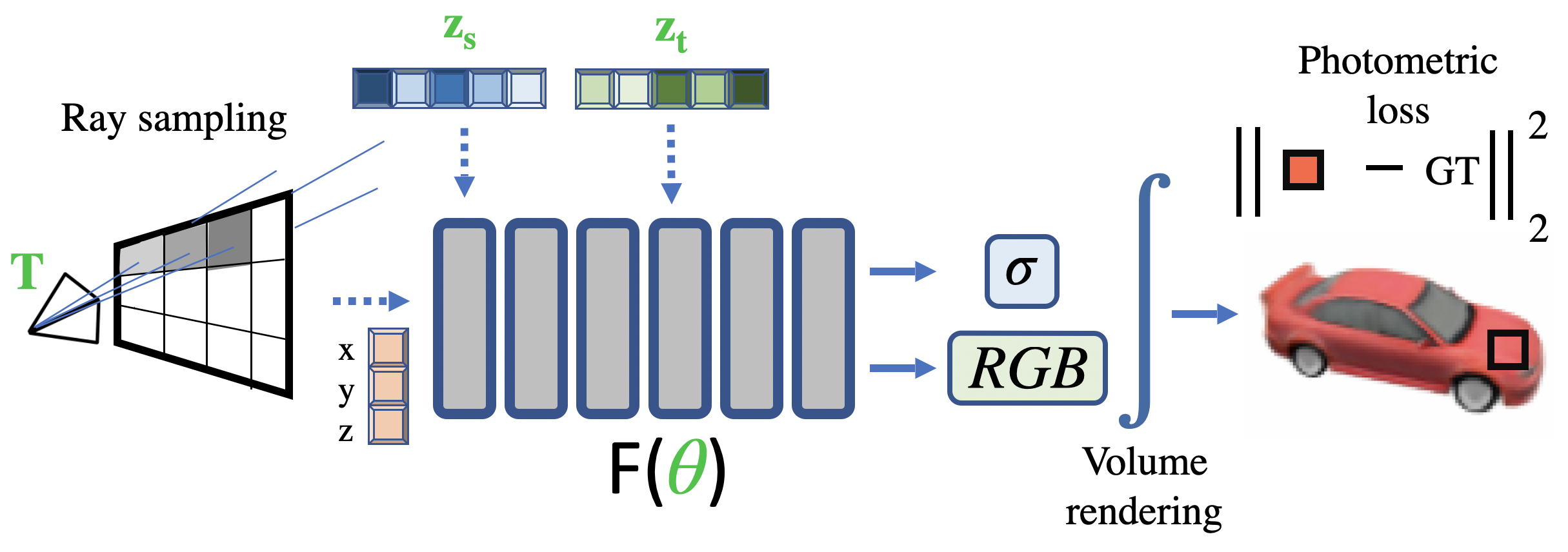}
\caption{CodeNeRF disentangles geometry, appearance and viewpoint by  learning separate embeddings for shape and texture and a fully connected network F($\theta$) that maps 3D locations and ray directions to density and RGB values.}
\label{fig:system}
\end{figure}

\begin{table}[t]
    \centering
    
\setlength{\tabcolsep}{3pt}
\begin{tabular}{@{}lccccc@{}}
\toprule
                    & \multicolumn{1}{c}{\begin{tabular}[c]{@{}c@{}}Deep\\ SDF\end{tabular}}      & SRN             & NeRF           & \multicolumn{1}{c}{\begin{tabular}[c]{@{}c@{}}Pixel\\ NeRF\end{tabular}} & CodeNeRF             \\ \midrule
Learned Prior &      \checkmark  &  \checkmark  & \xmark & \checkmark  & \checkmark    \\
Supervision         & 3D  & 2D & 2D & 2D  & 2D     \\
Encoder & \xmark & \xmark & \xmark & \checkmark & \xmark \\ 
Pose Optimization & - & \xmark & \xmark & \xmark & \checkmark \\ 
{\begin{tabular}[c]{@{}l@{}}Disentangled\\ Shape \& Texture \end{tabular}}      & \xmark           & \xmark & \xmark & \xmark & \checkmark            \\
\bottomrule
\\
\end{tabular}
    \caption{Related prior work on neural scene representations.}
    \label{tab:related_work_comparison}
    \vspace{-1em}
\end{table}

Neural representations have also been used to learn deformation priors that encode the variation of object shapes across semantic categories using direct 3D supervision~\cite{groueix2018papier,chen2019net,park2019deepsdf,mescheder2019occupancy}. DeepSDF~\cite{park2019deepsdf} is an auto-decoder architecture that jointly learns a latent embedding and the weights of a fully connected network that maps 3D coordinates to signed distance values. This representation uses test time optimization to estimate the shape code associated with a new unseen object. While DeepSDF~\cite{park2019deepsdf} has been an extremely  popular representation, successfully used as a shape prior to drive 3D reconstruction of object categories from multiple  images~\cite{runz2020frodo,remelli2020meshsdf}, its training requires 3D supervision and cannot be learnt from 2D images only. 

\begin{figure*}[t]
    \centering
    \includegraphics[width=0.95\textwidth]{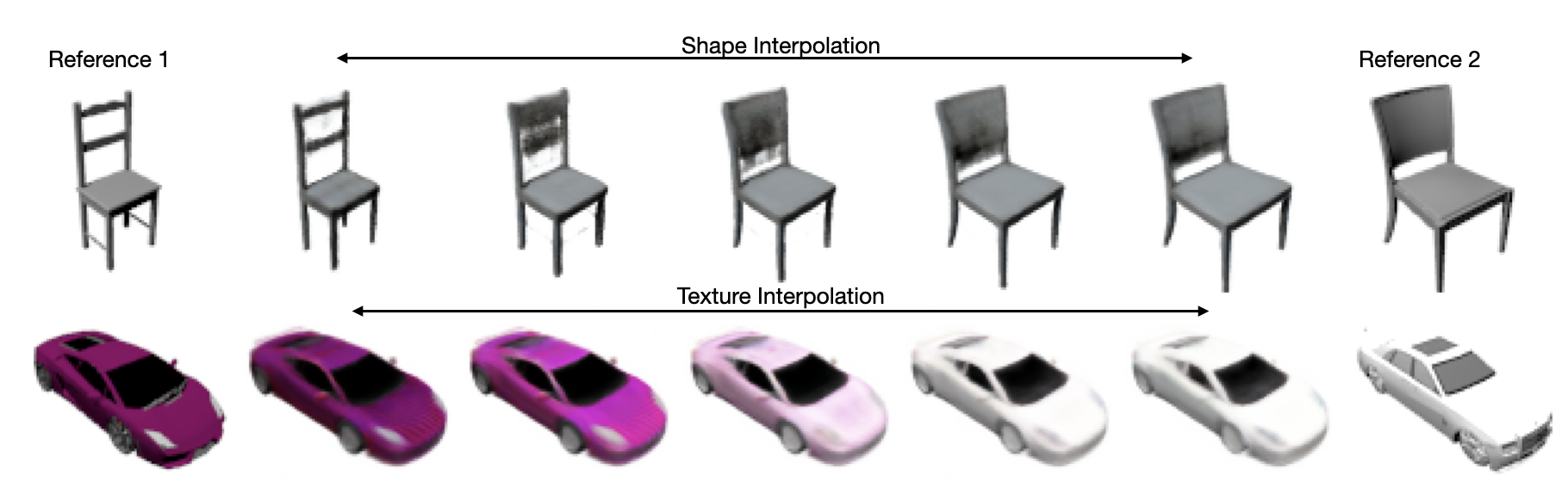}
        \vspace{0.1cm}
  \caption{We illustrate the ability of CodeNeRF to disentangle geometry and appearance by synthesizing new shapes or textures via latent interpolation. We show two reference images (left and right). CodeNeRF first estimates the camera pose and shape+texture codes for these images and then interpolates between them.} 
  \label{fig:novelshapetexture}
\end{figure*}

\noindent{\bf Contributions:} We present {\bf CodeNeRF}, a novel 3D-aware representation for object categories that learns to disentangle shape and texture. During training, CodeNeRF takes a set of posed input images and simultaneously learns different latent embeddings for shape and texture, and the weights of a multi-layer perceptron to predict volume density and view-dependent radiance for each 3D point by enforcing multi-view photometric consistency. At inference, given a single unposed reference image of an unseen object, CodeNeRF optimizes shape and texture codes as well as camera pose. Our disentangled representation provides full control over the synthesis task, enabling explicit editing of object shape and texture simply by modifying the respective latent codes (see Fig~\ref{fig:teaser}). We demonstrate single view reconstruction, novel view synthesis and shape/texture editing on the SRN benchmark and real-world images.

While our work takes inspiration from other continuous neural scene representations~\cite{mildenhall2020nerf,sitzmann2020scene}, it addresses many of their limitations. Unlike NeRF~\cite{mildenhall2020nerf}, CodeNeRF is not scene specific and can model the variations of shape and appearance across an object class. In contrast to SRN~\cite{sitzmann2020scene}, CodeNeRF disentangles geometry and appearance offering explicit control over shape and texture for synthesis tasks. Unlike both, CodeNeRF does not require knowledge of camera pose at test time, estimating it via optimization. Inspired by DeepSDF~\cite{park2019deepsdf}, we adopt an auto-decoder architecture but depart significantly as we only require 2D supervision and can disentangle shape and texture variations across an object category. See Fig~\ref{fig:system} for an overview of CodeNeRF.

\section{Related Work}

We focus our literature review on learning based approaches to 3D reconstruction and new view synthesis.

\noindent{\bf \text{3D} Reconstruction from \text{3D} supervision} Voxel-based methods were amongst the first~\cite{wu20153d, choy20163d, 3dgan} to propose learned representations for 3D reconstruction with 3D supervision. However, volumetric representations are fundamentally limited in terms of representation power and resolution and fail to capture surface details. Although point clouds are a common alternative to voxels, their unstructured nature and permutation-invariance make them  difficult to handle using cnn architectures. PointNet~\cite{qi2017pointnet} offered a unified  architecture able to deal with unordered point sets via maxpooling, that may be used for multiple downstream learning tasks. Mesh-based representations have also been used for 3D shape estimation given 3D supervision~\cite{groueix2018papier} or applying graph convolutions~\cite{wang2018pixel2mesh, gkioxari2020mesh}. However, these methods still fail to capture fine details. DeepSDF~\cite{park2019deepsdf} pioneered a new popular direction to represent 3D geometry via a continuous mapping from 3D points to an implicit representation of shape. DeepSDF jointly learns the weights of a fully connected network that maps 3D coordinates to signed distance values, and a latent embedding. While in principle DeepSDF can represent shapes at arbitrary resolutions without increasing its memory footprint, its training requires 3D supervision. Other concurrent approaches used continuous implicit occupancy networks~\cite{mescheder2019occupancy, chen2019net} but also require 3D annotations. A CNN-based encoder to extract local features to provide conditioning priors was used in~\cite{xu2019disn}. However, all these methods require heavy pre-processing of the training data to compute the SDF of query points. In addition, if a mesh is required during  optimization for rendering, marching cubes must be used, which is not easily differentiable~\cite{liao2018deep}. Mesh-based, SDF or occupancy representations pre-learnt with 3D supervision~\cite{mescheder2019occupancy, chen2019net, xu2019disn} can be combined with a CNN encoder to perform single view reconstruction. However, as noted in~\cite{tatarchenko2019single}, these models often perform recognition instead of 3D reconstruction. 

\noindent{\bf \text{3D} Reconstruction from \text{2D} supervision} The emergence of differentiable renderers~\cite{kato2018neural, li2018differentiable, ravi2020accelerating} allowed to obtain gradients through the rendering process, enabling them to be embedded into end-to-end learning frameworks. DeepSDF was used as a deep pre-learnt shape prior for 3D shape optimization from multiple images~\cite{runz2020frodo} requiring 2D supervision only. MeshSDF~\cite{remelli2020meshsdf} circumvents the non-differentiability of marching cubes and uses 2D supervision based on a silhouette loss. A mesh-based representation is optimized from 2D images in~\cite{lin2019photometric}. The disadvantage of many of these methods is that the learning of shape priors is decoupled from the shape estimation.

\noindent{\bf Neural 3D representations}  DVR~\cite{niemeyer2020differentiable} proposes a differentiable rendering formulation for implicit shape and texture representations that can be learnt from multi-view images and object masks only, without the need for 3D supervision. Sitzmann \etal~\cite{sitzmann2020scene} introduced SRN (Scene Representation Networks) a 3D-aware representation that learns a scene prior via a continuous mapping of 3D spatial locations to features of scene properties. Similar to our approach, once trained, the model can be used as a learned scene prior via test-time optimization. However, unlike CodeNeRF, test-time optimization for SRN requires known absolute camera poses, which can be seen as strong supervision. 

\noindent{\bf Neural Radiance Fields (NeRF) for new view synthesis} 
NeRF~\cite{mildenhall2020nerf} extended the neural representation to allow to capture viewpoint varying image properties by storing volume density and view-dependent radiance values which enabled photorealistic new view synthesis of complex real-world scenes. Many new extensions to NeRF have been recently proposed, some concurrent to ours. GRAF \cite{schwarz2020graf} introduces the idea of \emph{conditional} radiance fields into the generative adversarial network framework, to convert it into a 3D aware generative model. In addition to new viewpoint synthesis, similarly to ours, their approach allows to modify shape and appearance of the generated objects via shape and texture latent embeddings.  However, unlike ours, their method is a purely generative model, trained on an adversarial loss over patches. As such, it cannot estimate shape, texture or camera pose given an input image so cannot be used for single view reconstruction. PixelNeRF~\cite{yu2020pixelnerf} can generalise to multiple scenes by using an image encoder to condition the neural radiance field on image features and does not require test-time optimization. However, unlike our approach, there is no disentanglement of geometry and appearance and therefore no control over shape or texture editing. iNeRF~\cite{yen2020inerf} uses a pre-trained NeRF to optimize the camera pose while NeRF\texttt{-{}-}\cite{wang2021nerf} jointly estimates the neural representation and camera intrinsic and extrinsic parameters but both are scene specific.

\section{Methodology}
\begin{figure}[t]
    \centering
    \includegraphics[width=.48\textwidth]{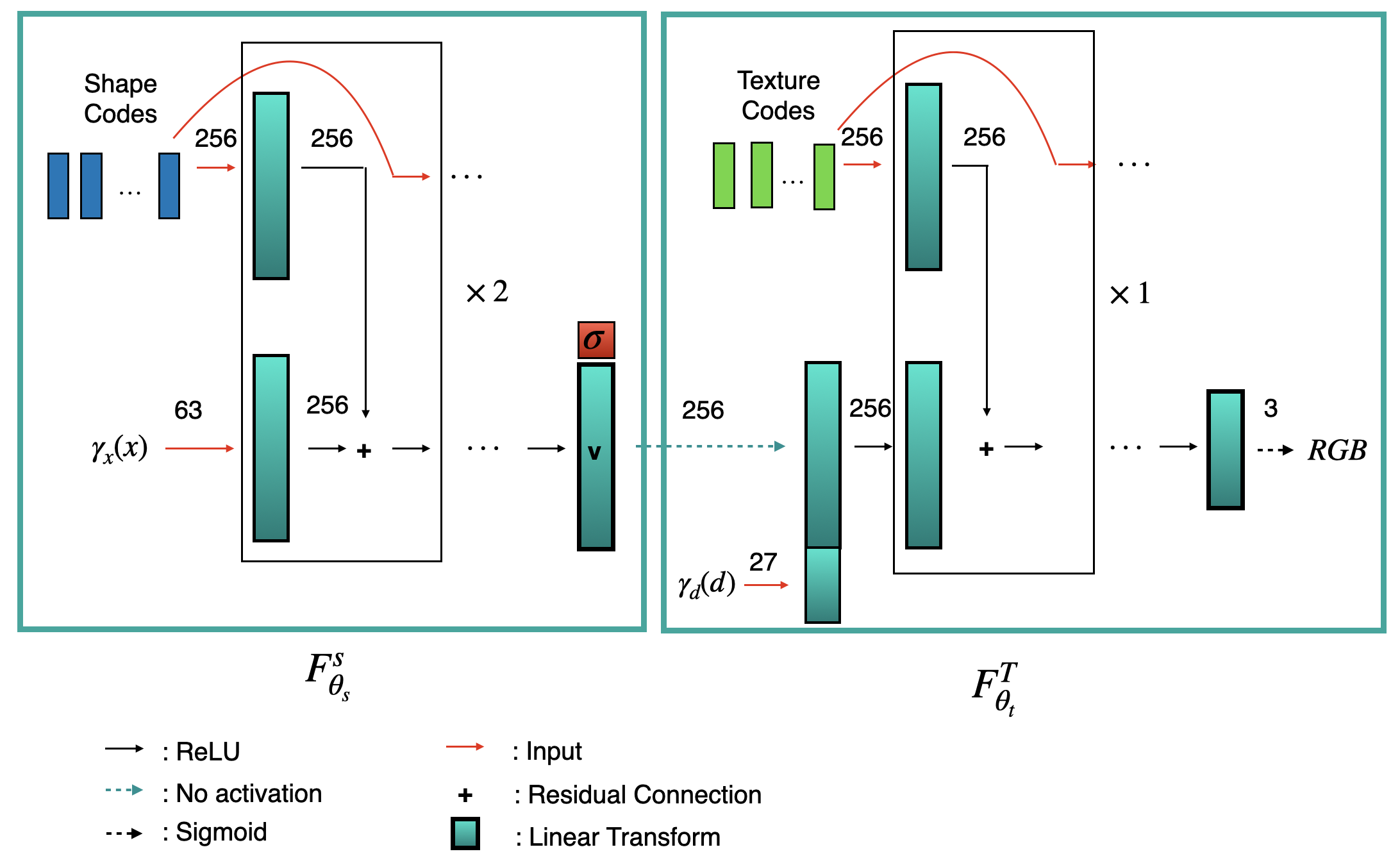}
    \caption{\textbf{CodeNeRF architecture.} $\gamma_x(\textbf{x})$ and  $\gamma_d(\textbf{d})$ are positional encodings for 3D point coordinates and viewing directions respectively. Volume density $\sigma$ does not depend on texture codes $\mathbf{z_t}$}
\label{fig:arch}
\end{figure}

Given a training set of $\text{N}$ images depicting $\text{M}$  objects across a semantic class, along with their respective camera intrinsics and pose parameters $\mathcal{V}=\{\mathcal{I}_i, \text{K}_i, \text{T}_i\}_{i=1}^\text{N}$, CodeNeRF jointly learns the weights of a multi-layer perceptron $\text{F}_\Theta$ that encapsulates the geometry and appearance across the observed objects, and separate latent embeddings $\{\mathbf{z_s}^j, \mathbf{z_t}^j\}_{j=1}^\text{M} \in \mathbb{R}^{256}$ that disentangle shape and appearance respectively. $\text{F}_\Theta$ is a disentangled neural radiance field that takes shape and texture codes as inputs and maps scene coordinates $\mathbf{x}$ and viewing directions $\mathbf{d}$ to their corresponding volume density $\sigma$ and $\text{RGB}$ colour values $\mathbf{c}$:

\begin{equation}
         \text{F}_\Theta :(\gamma_x(\mathbf{x}),\gamma_d(\mathbf{d}),\mathbf{z_s},\mathbf{z_t}) \xrightarrow[]{} (\sigma,\mathbf{c})
\label{eq:disentangled_radiance_field}
\end{equation}
Following~\cite{mildenhall2020nerf} we use positional encoding $\gamma(\cdot)$ for both scene coordinates $\mathbf{x}$ and viewing directions $\mathbf{d}$ to capture high frequency details. We employ $10$ frequencies for $\mathbf{x}$ and $4$ for $\mathbf{d}$. 
We design the architecture of the decoder to take advantage of the fact that the volume density $\sigma$ depends only on the 3D point $\mathbf{x}$ and shape code $\mathbf{z_s}$, while the \text{RGB} color depends in addition on the viewing direction $\mathbf{x}$ and texture code $\mathbf{z_t}$. The first layers of the MLP $\text{F}^{{s}}_\Theta$ map the input 3D coordinate $\gamma(\mathbf{x})$ and shape code $\mathbf{z_s}$ to the volume density $\sigma$ and an intermediate feature vector $\mathbf{v} \in \mathbb{R}^{256}$. The second part of the network $\text{F}^{{t}}_\Theta$ takes $\mathbf{v}$ and $\mathbf{z_t}$ as input and outputs the $\text{RGB}$ colour as shown in Fig.~\ref{fig:arch}. 
\begin{equation}\label{eq:detailedradiancefield}
\begin{split}
         \text{F}^{{s}}_{\Theta_s}& :(\gamma_{\text{x}}(\mathbf{x}),\mathbf{z_s}) \xrightarrow[]{} (\sigma,\mathbf{v})\\
         \text{F}^{{t}}_{\Theta_t}& :(\mathbf{v},\gamma_{\text{d}}(\mathbf{d}),\mathbf{z_t}) \xrightarrow[]{} (\mathbf{c})\\
         \text{F}_\Theta&:\text{F}^{{s}}_{\Theta_s} \circ \text{F}^{{t}}_{\Theta_t}
\end{split}
\end{equation}
\noindent{\bf Conditioning on Learned Shape and Texture Embeddings} Inspired by SRN~\cite{sitzmann2020scene} and DeepSDF~\cite{park2019deepsdf}, CodeNeRF adopts an auto-decoder architecture with decoupled shape and texture embedding spaces. During training, embedding vectors $\{\mathbf{z_s}^j,\mathbf{z_t}^j\}_{j=1}^\text{M}$ are optimized jointly with the parameters of the network $\text{F}_{\Theta}$. Unlike DeepSDF~\cite{park2019deepsdf} the shape and texture embeddings can be trained end-to-end from images only, without requiring any 3D supervision. In contrast to SRNs~\cite{sitzmann2020scene} shape and texture embeddings are decoupled and can be used as separate controls for editing/synthesis. Figs.~\ref{fig:teaser}, \ref{fig:novelshapetexture} and~\ref{fig:novelshapetextureexp} show how the embeddings can be used as conditioning to render smooth interpolations between different shapes/textures in the latent spaces. Jointly, the shape/texture embeddings and the decoder network $\text{F}_{\theta}$ act as learned priors that model shape deformations and texture variations across a semantic category. Once the model is trained, given a single input image, it can be used for one-shot reconstruction of an unseen object, via a test-time optimization that estimates the corresponding shape and texture codes $\{\mathbf{z_s}^j,\mathbf{z_t}^j\}$ and even the camera pose (see Sec.~\ref{sect:pose_estimation}). We show that CodeNeRF generalises well to render novel viewpoints, demonstrating that the learnt priors help to complete unobserved geometry and texture patterns (see Figs.~\ref{fig:realdataopt} and~\ref{fig:realdatanovel}).

\noindent{\bf Volume Rendering} We follow~\cite{mildenhall2020nerf} and use classical volumetric rendering techniques to render the color of image pixels by aggregating the color and the occupancy density at $\text{N}$ uniform samples from the near to far bounds for the camera ray  $r(t)=c+t\mathbf{d}$ traced through each pixel. The estimated color $C(r)$ of the pixel can be expressed as:
\begin{equation}
    \hat{C}(r) = \sum_{i=1}^\text{N} T_i(1-\exp(-\sigma_i\delta_i))\mathbf{c}_i
    \label{eq:nerf}
\vspace{-.1cm}
\end{equation}

where $\delta_i=t_{i+1}-t_i$ is the distance between adjacent samples and $T_i = \exp(-\sum_{j=1}^{i-1}\sigma_j \delta_j)$. This expression is fully differentiable, and reduces to the traditional alpha compositing with alpha values. NeRF~\cite{mildenhall2020nerf} adopts a 2 stage approach, a coarse and a fine network, with the fine network using importance sampling based on the values of $T_i$ from the coarse network to sample closer to the surface.

\subsection{Training CodeNeRF} 

For each image in the training set $\mathcal{V}=\{\mathcal{I}_i, \text{K}_i, \text{T}_i\}_{i=1}^\text{N}$, at each training iteration we sample a batch of 4094 rays using the intrinsic and extrinsic parameters. We then sample $64$ points along each ray and feed them to the MLP $F_{\Theta}$ together with the current estimates of shape and texture codes. $F_{\Theta}$ provides the occupancy density $\sigma$ and the RGB color $\mathbf{c}$ and volume rendering is used to aggregate colors along each ray. Training is supervised using the photometric loss along with a regularization loss used as a prior over the latent vectors (see Eq. $\eqref{eq:loss_train}$). We train the model with AdamW \cite{loshchilov2017decoupled} optimizer with initial learning rates 1e-4 for the network parameters and 1e-3 for the latent vectors.

\begin{equation}
\begin{split}
\mathcal{L}&(\Theta, \{\mathbf{z_s}^i, \mathbf{z_t}^i\}^\text{M}) =\sum_{r \in \mathcal{R}} ||\hat{C(r)} - C(r)||_2^2 \\
\min_{\Theta,\{\mathbf{z_s}^i, \mathbf{z_t}^i\}^\text{M}} \mathcal{L}&(\Theta, \{\mathbf{z_s}^i, \mathbf{z_t}^i\})  + \frac{1}{\nu^2} (||\mathbf{z_s}^i||^2_2+||\mathbf{z_t}^i||^2_2) 
\end{split}   
\label{eq:loss_train}
\end{equation}

\subsection{Inference Optimization} 
\label{sect:pose_estimation}

At inference time, given a single input image of an unseen object, the trained model can be used to optimize the latent vectors by minimizing the photometric loss between the rendered and observed pixels (Eq.~\ref{eq:loss_inf}), while keeping the weights of the neural network fixed. Moreover, unlike previous methods~\cite{sitzmann2020scene,yu2020pixelnerf} we show that CodeNeRF does not require known camera pose and its parameters can also be optimized jointly with the latent codes.

\noindent{\bf Camera Pose Optimization:} CodeNeRF backprojects pixels into 3D rays using the camera to world transformation matrix. Similarly to NeRF~\cite{mildenhall2020nerf}, we require known camera matrices to train the model. However, once the model has been learnt, at inference time, unlike others~\cite{sitzmann2020scene,mildenhall2020nerf,yu2020pixelnerf}, we do not require known camera viewpoint and we optimize the rotation and translation parameters jointly with the latent embedding vectors, by obtaining the gradients over the input points and ray directions, in a similar way to a differentiable renderer.
\begin{figure}[t]
    \centering
    \includegraphics[width=0.7\linewidth]{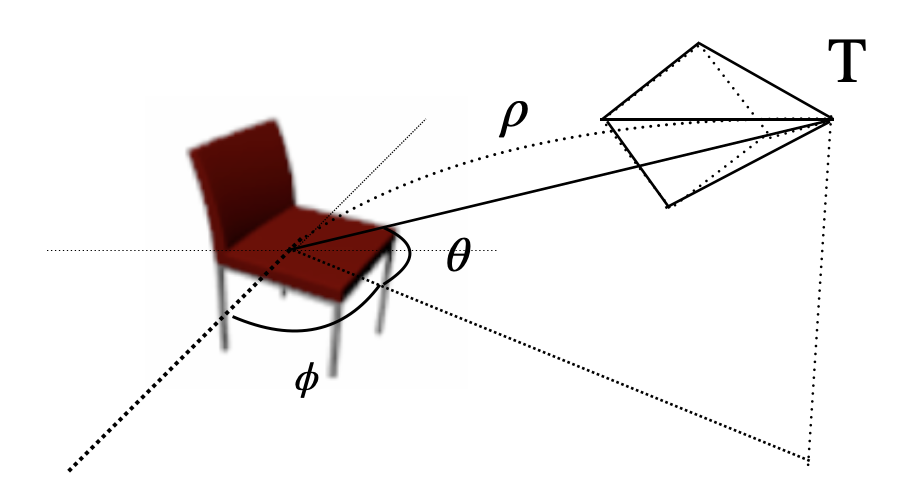}
    \caption{Camera viewpoint parametrization. At test time, CodeNeRF optimizes camera pose jointly with latent codes.}
\label{fig:cam}
\end{figure}
Similarly to other object-category 3D reconstruction methods, we assume a simplified camera model, where the world coordinate system is centred on the 3D object, the camera faces its origin, and the up vector is aligned with the $z$ axis. In this case, $\textbf{R}$ and $\textbf{t}$ can be parametrized via azimuth $\phi$, elevation $\theta$ and distance to the camera $\rho$, as shown in Fig.~\ref{fig:cam}. The camera to world transformation matrix can be written as $\textbf{T}_{cw} = \big( \begin{smallmatrix} \textbf{R}^T & \textbf{p}\\ 0 & 1 \end{smallmatrix} \big)$ where:
\begin{equation}\nonumber
\textbf{R} = 
\begin{pmatrix}
-\sin\phi & \cos\phi & 0 \\
-\sin\theta\cos\phi & -\sin\theta\sin\phi & \cos\theta \\
\cos\theta\cos\phi & \cos\theta\sin\phi & \sin\theta \\
\end{pmatrix}
\label{eq:camrot}
\end{equation}
and $\textbf{p}=(\rho\cos\theta\cos\phi, \rho\cos\theta\sin\phi, \rho\sin\theta)$ is the camera location. At inference time, given an reference image $\mathcal{I}_i$, the camera pose and the sampled rays $\gamma_{i}$ can be expressed in terms of the camera parameters azimuth $(\phi, \theta, \rho)$ and optimized via using the gradients. 

\noindent{\bf Test time optimization:} We minimize the photometric loss $\eqref{eq:loss_inf}$ jointly with respect to shape and texture codes and camera parameters (fixing the decoder parameters $\Theta$), using the AdamW optimizer. 
\begin{equation}
    \min_{\mathbf{z_s}^i,\mathbf{z_t}^i,\rho_i,\theta_i,\phi_i} \mathcal{L}(\mathbf{z_s}^i,\mathbf{z_t}^i,\rho_i,\theta_i,\phi_i) + \frac{1}{\nu^2} (||\mathbf{z_s}^i||^2_2+||\mathbf{z_t}^i||^2_2) 
\label{eq:loss_inf}
\end{equation}
Latent vectors are initialized with the mean vector of the trained embeddings, and an initial learning rate of 1e-2 is used. The initial learning rates for the camera parameters are 1e-2, 1e-1 and 1e-1 respectively.

\begin{figure}[h]
    \centering
    \includegraphics[width=0.9\linewidth]{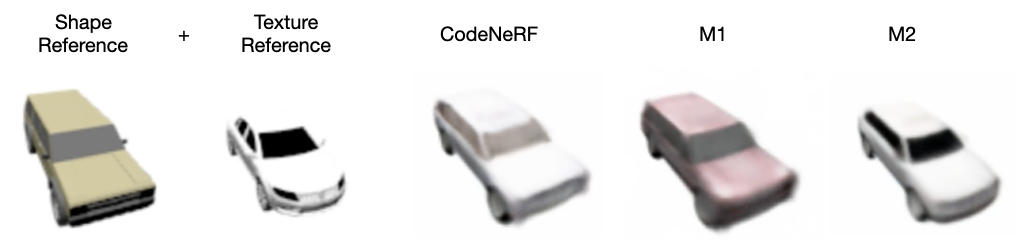}
    \caption{Shape/Texture disentanglement: Alternative architectures fail to disentangle shape and texture.}
\label{fig:disentangle}
\end{figure}

\noindent{\bf Enforcing disentanglement:}
To illustrate the link between good architecture choices and disentanglement, we trained two alternative models (M1/M2). In M1, $\gamma_d(d)$ is given as input to the first layer of $F^s_{\theta_s}$ jointly with 3D point position $\gamma_x(x)$ (similarly to PixelNeRF). In M2, $z_s$ and $z_t$ are given as input to $F^s_{\theta_s}$, which amounts to using a single embedding.  We then perform texture editing keeping the shape code fixed. Fig.~\ref{fig:disentangle} shows how only CodeNeRF fully disambiguates shape and texture. M1 fails to synthesise the
correct texture while M2 has unwanted changes in shape.

\section{Experimental Evaluation}

\begin{table}[t]
    \centering

\setlength{\tabcolsep}{5.0pt}
\begin{tabular}{@{}lllll@{}}
                    &                            \multicolumn{2}{c}{1-view}                                                &                       \multicolumn{2}{c}{2-view}                                                \\ \cmidrule(lr){2-3} \cmidrule(l){4-5} 
                    &   \multicolumn{1}{c}{PSNR} & \multicolumn{1}{c}{SSIM} & \multicolumn{1}{c}{PSNR} & \multicolumn{1}{c}{SSIM} \\ \midrule
Chairs & & & & \\ \midrule

GRF~\cite{trevithick2020grf}  & 21.25 & 0.86 & 22.65 & 0.88 \\
TCO~\cite{tatarchenko2015single} & 21.27  & 0.88 & 21.33 & 0.88 \\
dGQN~\cite{eslami2018neural} & 21.59 & 0.87 & 22.36  & 0.89 \\
ENR~\cite{dupont2020equivariant}& 22.83& - & -& -  \\
SRN~\cite{sitzmann2020scene} & 22.89 & 0.89 & 24.48 & 0.92  \\
PixelNeRF~\cite{yu2020pixelnerf} & \bf{23.72} & \bf{0.91} & \bf{26.20} & \bf{0.94} \\ 
CodeNeRF (GT pose) & 23.66 & 0.90 & 25.63 & 0.91 \\
CodeNeRF  & 22.39 & 0.87 & - & - \\ 
CodeNeRF ($-$ outliers) & 23.11 & 0.89 & - & - \\\midrule
Cars & & & & \\ \midrule
SRN~\cite{sitzmann2020scene} & 22.25 & 0.89 & 24.84  & 0.92\\
ENR~\cite{dupont2020equivariant} & 22.26 & - & -  & - \\
PixelNeRF~\cite{yu2020pixelnerf} & 23.17 & 0.90 & 25.66 & \bf{0.94} \\ 
CodeNeRF (GT pose) & \bf{23.80} & \bf{0.91} &  \bf{25.71} & 0.93 \\
CodeNeRF & 22.73 & 0.89 & - & - \\
CodeNeRF ($-$ outliers)& 23.17 & 0.90 & - & - \\
                    \bottomrule
\end{tabular}

    \caption{Quantitative evaluation on ShapeNet-SRN}
    \label{tab:compare_psnr_ssim}
    \vspace{-1em}
\end{table}

\begin{figure}[t]
    \centering
    \includegraphics[width=\linewidth]{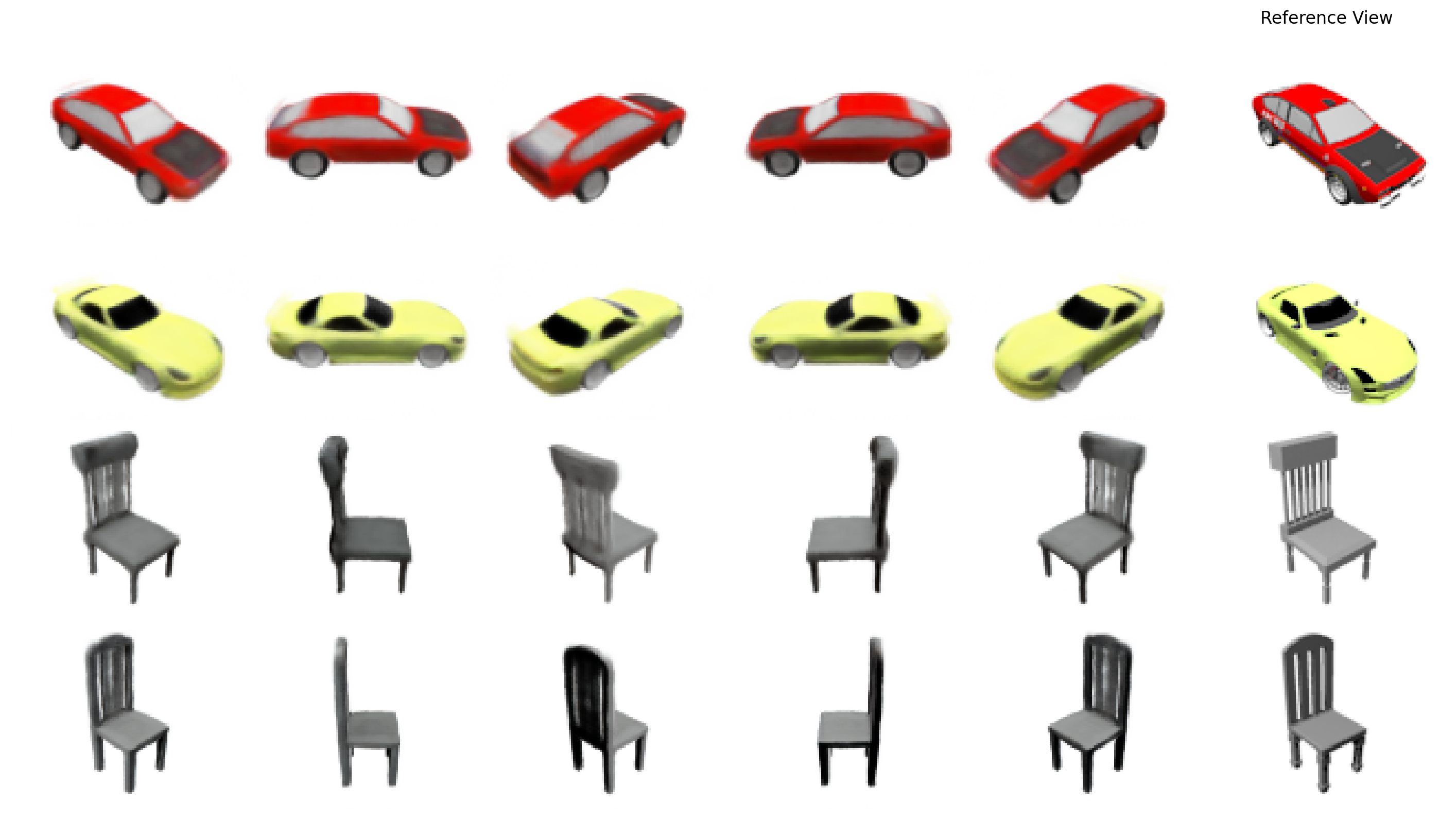}
      \caption{Qualitative results of novel view synthesis of unseen objects from a single input image with CodeNeRF (GT pose), a variant of CodeNeRF which assumes known camera pose at test time. Rightmost column shows input view. }
\label{fig:novelviewsynthesisunknown}
\end{figure}

\begin{figure}[t]
    \centering
    \includegraphics[width=0.95\linewidth]{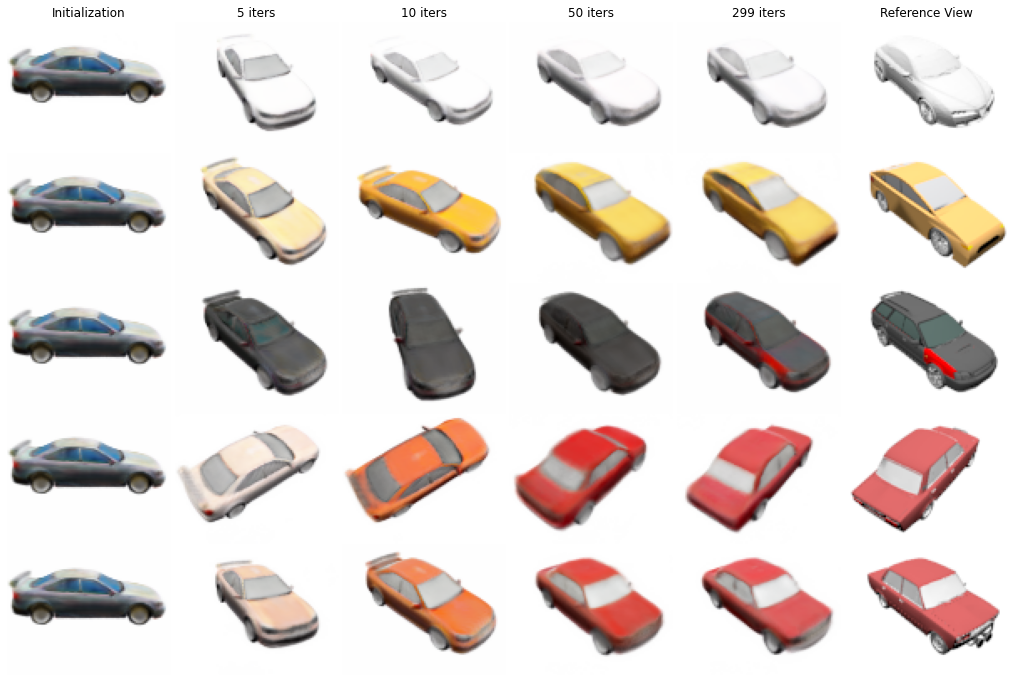}
        \vspace{0.1cm}
  \caption{Test-time joint optimization of camera pose and latent codes: Evolution of  estimated renderings from initialization (left), through intermediate optimization  iterations $5/10/50$, to final result after $299$ iterations.}
  \label{fig:optpose}
\end{figure}

\begin{figure*}[t]
    \centering
    \includegraphics[width=0.95\textwidth]{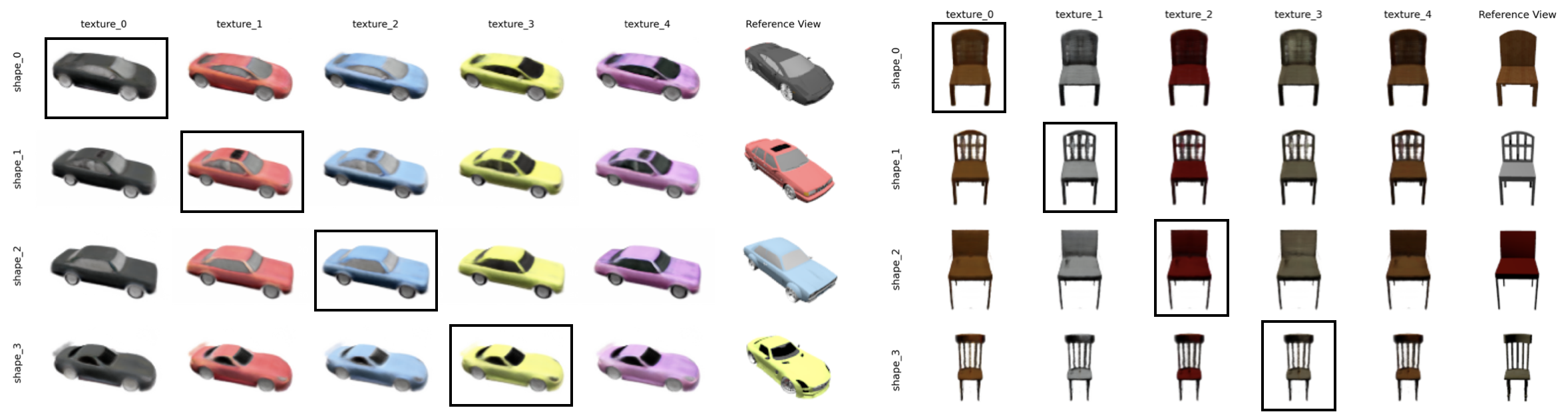}
        \vspace{0.1cm}
  \caption{Novel Shape/Texture/Pose Synthesis. Highlighted results show the renderings with shape and texture codes corresponding to the reference views (obtained via optimization). Others show results of varying shapes and textures simply by editing corresponding latent codes. } 
  \label{fig:novelshapetextureexp}
\end{figure*}

\begin{figure}[t]
    \centering
    \includegraphics[width=0.95\linewidth]{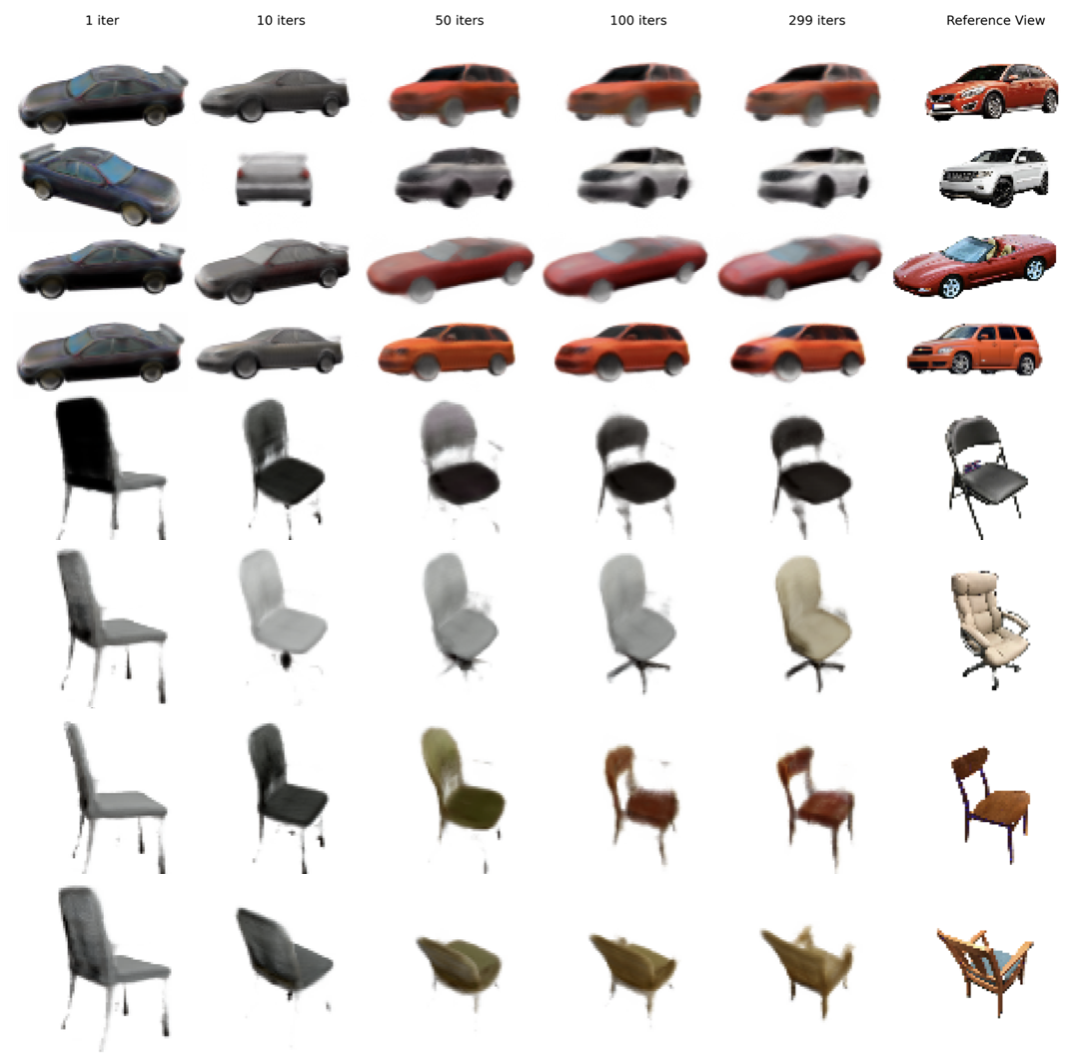}
        \vspace{0.1cm}
  \caption{Qualitative results on real-world datasets (Stanford-Car~\cite{KrauseStarkDengFei-Fei_3DRR2013} and Pix3D~\cite{pix3d}) of CodeNeRF with joint optimization of camera pose+latent codes. Evolution of estimated renderings throughout the optimization from initialization through iterations $5/10/50$ and final result after $299$ iterations. Rightmost column shows input image.}
  \label{fig:realdataopt}
\end{figure}
We train our model on the ShapeNet-SRN dataset of ShapeNet~\cite{chang2015shapenet} renderings for two object categories: cars (3514) and chairs (6591), created by Sitzmann \etal~\cite{sitzmann2020scene}. We train a different model for each category using the pre-defined train/test split and conduct a  quantitative evaluation on novel view synthesis and few shot reconstruction tasks.

\subsection{Quantitative Evaluation on SRN Benchmark}

The ShapeNet-SRN benchmark~\cite{sitzmann2020scene} provides 251 test images on an Archimedeal spiral per object in the test set. We follow the evaluation protocols provided by SRN~\cite{sitzmann2020scene}. 

\noindent{\bf Baselines} We compare quantitatively on the ShapeNet benchmark~\cite{sitzmann2020scene} with  SRN~\cite{sitzmann2020scene}, ENR~\cite{dupont2020equivariant}, PixelNeRF~\cite{yu2020pixelnerf}, and other current state of the art methods, on the tasks of one-view and two-view 2D-supervised reconstruction. Note that competing approaches require known camera pose at inference time. We compare with three variants of our approach: {CodeNeRF (GT pose)} assumes known camera pose at test time, {CodeNeRF} is our full approach with joint optimization of camera pose and shape/texture codes (299 iterations) and {CodeNeRF ($-$ outliers)}, which is CodeNeRF but without counting results that resulted in an outlier (rotation error above $5^\circ{}$ or translation error higer than $3\%$).

\noindent{\bf Metrics} We use PSNR(Peak Signal to Noise Ratio) and SSIM(Structural Similarity Index Measure)~\cite{wang2004image} to evaluate the quality of the rendered images. 

\noindent{\bf Synthesis Results} Table \ref{tab:compare_psnr_ssim}, shows a full comparison of the results on 1-view and 2-view reconstruction. CodeNeRF (GT pose) -- which assumes known camera pose, for a fairer comparison with its competitors that need it as input --  outperforms PixelNeRF for cars and comes very close on chairs. Even when CodeNeRF estimates the unknown pose at test time, along with the latent codes, the performance does not degrade much. When pose estimation outliers are removed the performance is close to PixelNeRF. Qualitative results for 1-view reconstruction with CodeNeRF (GT pose) are shown in Fig.~\ref{fig:novelviewsynthesisunknown} while  Fig.~\ref{fig:optpose} shows results for CodeNeRF with camera pose+latent codes optimization. Clearly, rendered images become sharper when the ground truth pose is used, but as Fig.~\ref{fig:optpose} shows, even when pose and latent codes are initialized far from the ground truth, CodeNeRF converges to good estimates. 

\noindent{\bf Editing Novel Shapes and Textures} After the test-time optimization has taken place, CodeNeRF provides full control over the synthesis process as object shapes and textures can be edited simply by varying the corresponding latent codes. This is possible given that viewpoint, shape and texture are fully disentangled in our representation. As shown in Fig.~\ref{fig:novelshapetextureexp}, we can easily fix one while changing the other. 

\begin{figure}[ht]
    \centering
    \includegraphics[width=0.95\linewidth]{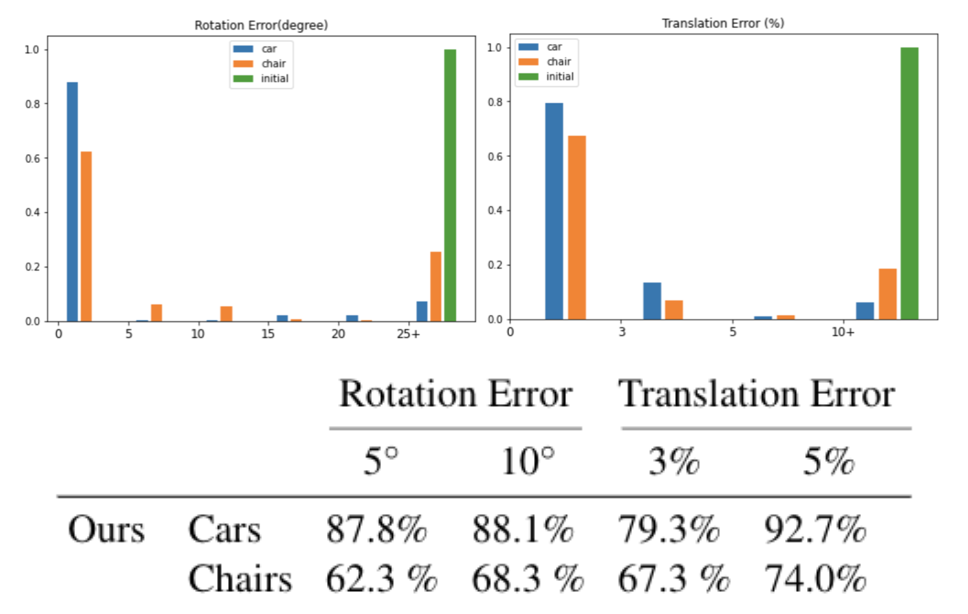}
        \vspace{0.1cm}
  \caption{Pose estimation evaluation. Error distribution histograms: rotation (top-left), translation (top-right). Initialization error in green. Numerical evaluation (below).}
  \label{fig:diffpose}
\end{figure}

\noindent{\bf Evaluation of Camera Viewpoint Estimation:} At test time, CodeNeRF optimizes jointly the camera pose and shape/pose latent codes. We run a quantitative evaluation of the performance of the camera pose estimation on the ShapeNet-SRN dataset. Fig.~\ref{fig:diffpose} (top) shows histograms with the distribution of pose estimation errors. It is interesting to note that the initializations we used for the camera pose parameters (shown in green) had rotation errors above $30^\circ{}$ in $100\%$ of cases and were on average $80^\circ{}$ away from GT. Despite this, our optimization converged in $85\%$ of cases. The table in Fig.~\ref{fig:diffpose} (below) shows the numerical errors between the estimated and ground-truth camera poses in terms of the percentage of rotation estimates with errors below ${5^\circ}$ and ${10^\circ}$, and the percentage of translation estimates with relative errors below ${3\%}$ and ${5\%}$.  
Fig.~\ref{fig:optpose} shows the initialization and illustrates the evolution of the intermediate results obtained through the optimization (iterations 5/10/50), and final result after 299 iterations. Even when the initial estimate is far away from the ground truth, the optimization converges to good solutions.

\subsection{Qualitative Comparisons}

\begin{figure}[ht]
    \centering
    \includegraphics[width=\linewidth]{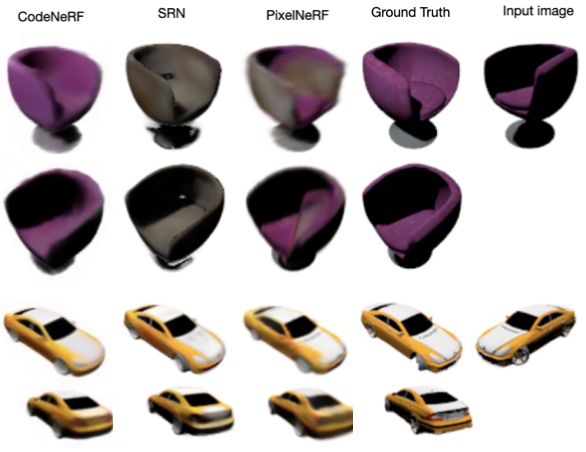}
      \caption{Qualitative comparisons with SRN~\cite{sitzmann2020scene} and PixelNeRF~\cite{yu2020pixelnerf} (results courtesy of A.~Yu and V.~Sitzmann).}
\label{fig:comparemodels}
\end{figure}

Fig.~\ref{fig:comparemodels} shows examples comparing CodeNeRF with PixelNeRF~\cite{yu2020pixelnerf} and SRN~\cite{sitzmann2020scene} on new view synthesis of objects from the ShapeNet-SRN test set. Since PixelNeRF uses a pre-trained CNN to extract features, it renders sharper images when the target view is close to the input view. SRN works better on occluded regions than PixelNeRF as it learns a latent embedding that acts as a priors. CodeNeRF achieves shape and texture disentanglement maintaining geometric and appearance consistency in occluded regions. Only CodeNeRF can synthesise the correct combination of shape and texture for the purple chair.

\subsection{Results on Real World Datasets}

To understand if our model, trained only on synthetic renderings from ShapeNet-SRN, can generalise well to real-world images, we perform an evaluation on two datasets: Stanford-Car dataset~\cite{KrauseStarkDengFei-Fei_3DRR2013} and Pix3D~\cite{pix3d}. We perform test time optimization given a single input image to estimate jointly camera pose+latent codes. We then render the representation from unseen viewpoints.

\noindent{\bf Pre-processing} Similarly to  PixelNeRF~\cite{yu2020pixelnerf}, we use Detectron2~\cite{wu2019detectron2} on the Stanford-Car dataset to infer masks, then apply Gaussian blur and downscale the images to $128\times128$. For the real chairs in Pix3D~\cite{pix3d}, we carve out the object with the provided ground-truth mask before downscaling.

\noindent{\bf Optimization of Camera Pose and Latent Vectors} We show results of one-shot reconstruction given a single input image with unknown camera pose. Since the camera intrinsics are unknown, we assume the same focal length as in ShapeNet-SRN. Fig.~$\ref{fig:realdataopt}$ shows the initialization and the intermediate results through different iterations of the optimization. The optimization provides convincing final renderings even when initialised from far away.

\begin{figure}[t]
    \centering
    \includegraphics[width=0.95\linewidth]{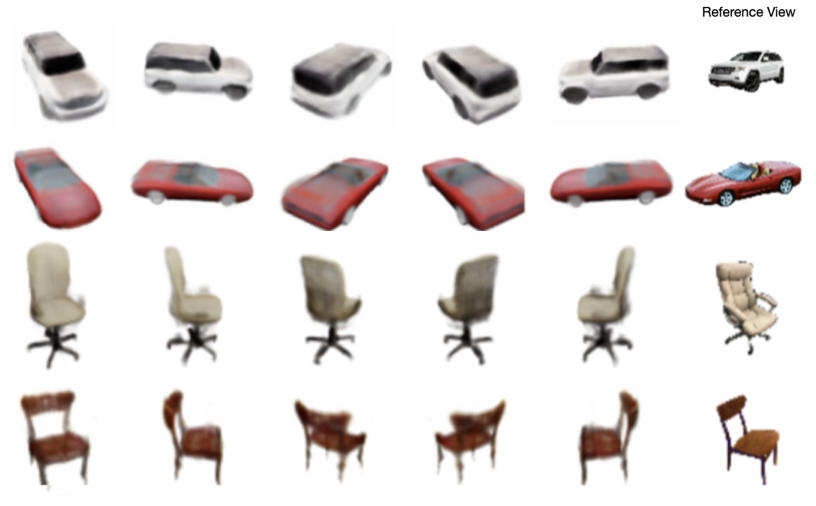}
        \vspace{0.1cm}
  \caption{Qualitative evaluation of novel view synthesis on images from real-world datasets (Stanford-Car~\cite{KrauseStarkDengFei-Fei_3DRR2013} and Pix3D~\cite{pix3d}). Input image is shown on the rightmost column. Camera pose is estimated jointly with  latent codes.}
  \label{fig:realdatanovel}
\end{figure}

\begin{figure}[t]
    \centering
    \includegraphics[width=0.95\linewidth]{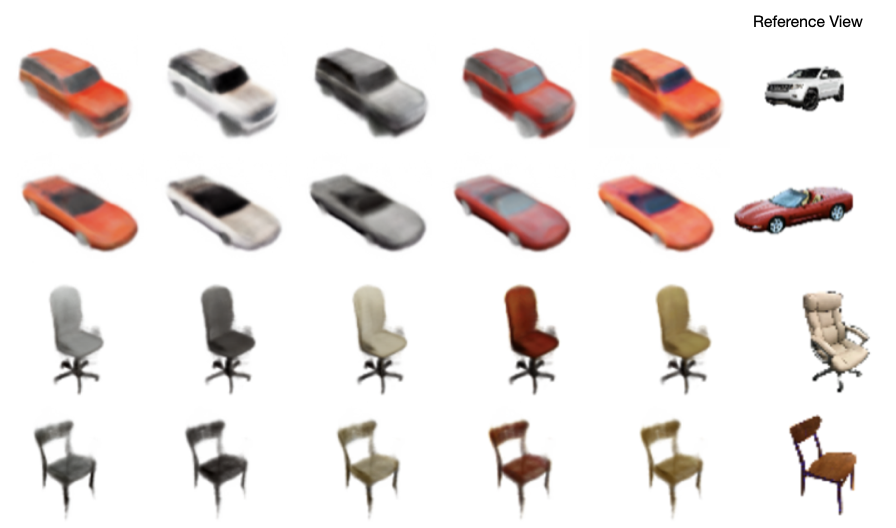}
        \vspace{0.1cm}
  \caption{{Shape and texture editing on real images from the Stanford-Cars and Pix3D datasets.}}
  \label{fig:realdataedit}
\end{figure}

\noindent{\bf Shape, Texture and Viewpoint Editing} Following  one-shot optimization, we show renderings of the reconstructed objects from novel viewpoints (see Fig.~\ref{fig:realdatanovel}). Fig.~\ref{fig:realdataedit} illustrates results after editing shape and texture latent vectors. The distinctive style of each object is preserved after shape or texture transfer. The shape and texture priors learned by CodeNeRF help to complete object shape and texture information that was not present in the input image. 

\begin{figure}[ht]
    \centering
    \includegraphics[width=0.95\linewidth]{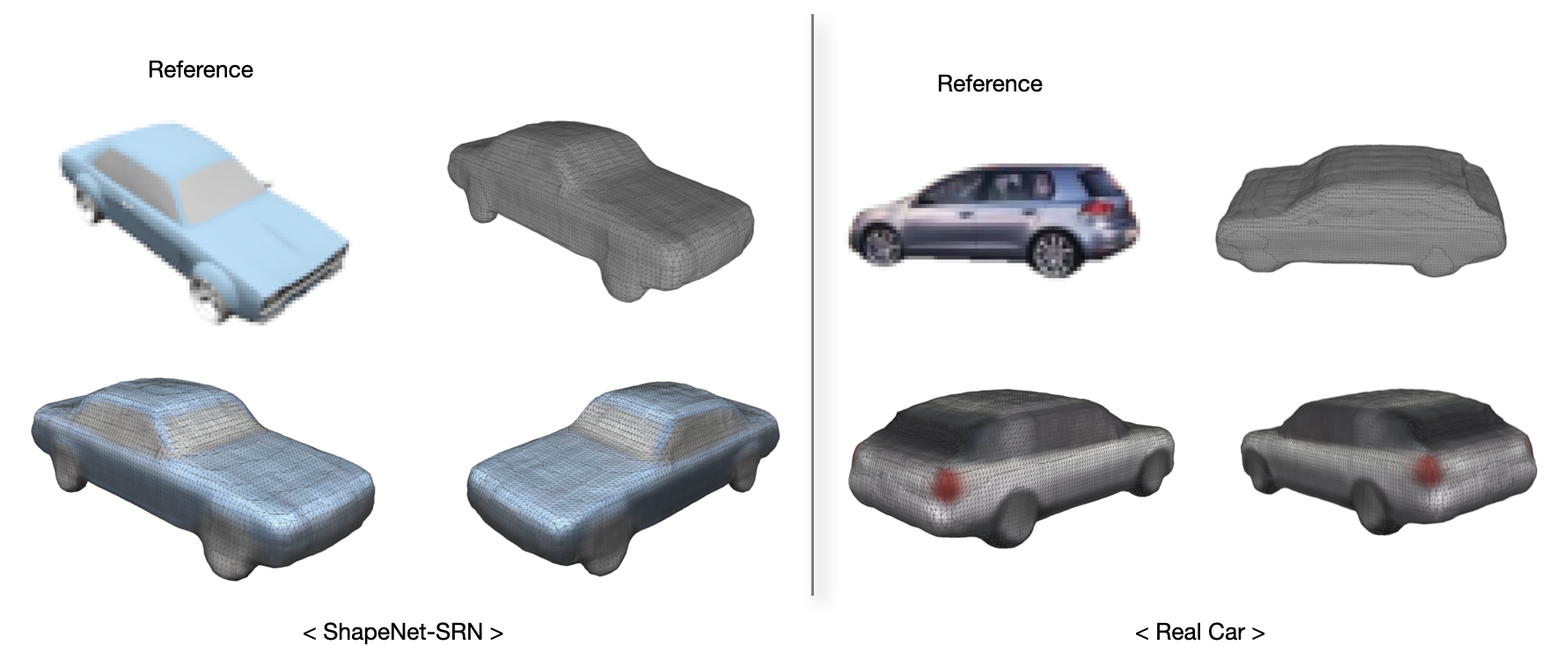}
        \vspace{0.1cm}
  \caption{\textbf{Reconstructing 3D meshes:} CodeNeRF allows 3D reconstruction of a mesh from a single input image.} 
  \label{fig:meshes}
\end{figure}

\noindent{\bf Explicit mesh reconstruction}
CodeNeRF implicitly represents the 3D structure of reconstructed objects. To obtain an explicit mesh representation, we feed the centre of of 256$^3$ voxels to the MLP, and obtain their volume density values $\sigma$. Marching cubes can then be used to find the mesh vertices. Casting rays along vertex normals we can associating colors to the vertices. 
Fig.~$\ref{fig:meshes}$ shows the input image and the reconstructed meshes for a real image from the Stanford-Car dataset and a synthetic car from ShapeNet.

\section{Conclusion}

We have presented CodeNeRF, a neural radiance field that learns to disentangle shape and appearance by jointly learning separate latent embeddings for shape and texture, and an MLP that maps continuous input locations and ray directions to density and colour information. At inference time, given a single input image CodeNeRF can estimate its associated camera pose and latent codes. We show extensive results of one-shot reconstruction on the SRN benchmark and on real world images (Stanford-Cars and Pix3D) that demonstrate that CodeNeRF learns powerful scene priors that enable full control over the synthesis process. 

\noindent{\bf Acknowledgements:} Research presented here has been supported by funding from Cisco to the UCL AI Centre.

{\small
\bibliographystyle{ieee_fullname}
\bibliography{egbib}

\begin{thebibliography}{10}\itemsep=-1pt

\bibitem{chang2015shapenet}
Angel~X Chang, Thomas Funkhouser, Leonidas Guibas, Pat Hanrahan, Qixing Huang,
  Zimo Li, Silvio Savarese, Manolis Savva, Shuran Song, Hao Su, et~al.
\newblock Shapenet: An information-rich 3d model repository.
\newblock {\em arXiv preprint arXiv:1512.03012}, 2015.

\bibitem{chen2019net}
Zhiqin Chen.
\newblock {\em IM-NET: Learning implicit fields for generative shape modeling}.
\newblock PhD thesis, Applied Sciences: School of Computing Science, 2019.

\bibitem{choy20163d}
Christopher~B Choy, Danfei Xu, JunYoung Gwak, Kevin Chen, and Silvio Savarese.
\newblock 3d-r2n2: A unified approach for single and multi-view 3d object
  reconstruction.
\newblock In {\em European conference on computer vision}, pages 628--644.
  Springer, 2016.

\bibitem{dupont2020equivariant}
Emilien Dupont, Miguel~Bautista Martin, Alex Colburn, Aditya Sankar, Josh
  Susskind, and Qi Shan.
\newblock Equivariant neural rendering.
\newblock In {\em International Conference on Machine Learning}, pages
  2761--2770. PMLR, 2020.

\bibitem{eslami2018neural}
SM~Ali Eslami, Danilo~Jimenez Rezende, Frederic Besse, Fabio Viola, Ari~S
  Morcos, Marta Garnelo, Avraham Ruderman, Andrei~A Rusu, Ivo Danihelka, Karol
  Gregor, et~al.
\newblock Neural scene representation and rendering.
\newblock {\em Science}, 360(6394):1204--1210, 2018.

\bibitem{gkioxari2020mesh}
Georgia Gkioxari, Jitendra Malik, and Justin Johnson.
\newblock Mesh r-cnn.
\newblock {\em 2019 IEEE/CVF International Conference on Computer Vision
  (ICCV)}, pages 9784--9794, 2019.

\bibitem{groueix2018papier}
Thibault Groueix, Matthew Fisher, Vladimir~G Kim, Bryan~C Russell, and Mathieu
  Aubry.
\newblock A papier-m{\^a}ch{\'e} approach to learning 3d surface generation.
\newblock In {\em Proceedings of the IEEE conference on computer vision and
  pattern recognition}, pages 216--224, 2018.

\bibitem{kato2018neural}
Hiroharu Kato, Yoshitaka Ushiku, and Tatsuya Harada.
\newblock Neural 3d mesh renderer.
\newblock In {\em IEEE Conference on Computer Vision and Pattern Recognition
  (CVPR)}, 2018.

\bibitem{KrauseStarkDengFei-Fei_3DRR2013}
Jonathan Krause, Michael Stark, Jia Deng, and Li Fei-Fei.
\newblock 3d object representations for fine-grained categorization.
\newblock In {\em 4th International IEEE Workshop on 3D Representation and
  Recognition (3dRR-13)}, Sydney, Australia, 2013.

\bibitem{li2018differentiable}
Tzu-Mao Li, Miika Aittala, Fr{\'e}do Durand, and Jaakko Lehtinen.
\newblock Differentiable monte carlo ray tracing through edge sampling.
\newblock {\em ACM Transactions on Graphics (TOG)}, 37(6):1--11, 2018.

\bibitem{liao2018deep}
Yiyi Liao, Simon Donne, and Andreas Geiger.
\newblock Deep marching cubes: Learning explicit surface representations.
\newblock In {\em Proceedings of the IEEE Conference on Computer Vision and
  Pattern Recognition}, pages 2916--2925, 2018.

\bibitem{lin2019photometric}
Chen-Hsuan Lin, Oliver Wang, Bryan~C Russell, Eli Shechtman, Vladimir~G Kim,
  Matthew Fisher, and Simon Lucey.
\newblock Photometric mesh optimization for video-aligned 3d object
  reconstruction.
\newblock In {\em Proceedings of the IEEE/CVF Conference on Computer Vision and
  Pattern Recognition}, pages 969--978, 2019.

\bibitem{loshchilov2017decoupled}
Ilya Loshchilov and Frank Hutter.
\newblock Decoupled weight decay regularization.
\newblock {\em arXiv preprint arXiv:1711.05101}, 2017.

\bibitem{mescheder2019occupancy}
Lars Mescheder, Michael Oechsle, Michael Niemeyer, Sebastian Nowozin, and
  Andreas Geiger.
\newblock Occupancy networks: Learning 3d reconstruction in function space.
\newblock In {\em Proceedings of the IEEE/CVF Conference on Computer Vision and
  Pattern Recognition}, pages 4460--4470, 2019.

\bibitem{mildenhall2020nerf}
Ben Mildenhall, Pratul~P Srinivasan, Matthew Tancik, Jonathan~T Barron, Ravi
  Ramamoorthi, and Ren Ng.
\newblock Nerf: Representing scenes as neural radiance fields for view
  synthesis.
\newblock {\em arXiv preprint arXiv:2003.08934}, 2020.

\bibitem{niemeyer2020differentiable}
Michael Niemeyer, Lars Mescheder, Michael Oechsle, and Andreas Geiger.
\newblock Differentiable volumetric rendering: Learning implicit 3d
  representations without 3d supervision.
\newblock In {\em Proceedings of the IEEE/CVF Conference on Computer Vision and
  Pattern Recognition}, pages 3504--3515, 2020.

\bibitem{park2019deepsdf}
Jeong~Joon Park, Peter Florence, Julian Straub, Richard Newcombe, and Steven
  Lovegrove.
\newblock Deepsdf: Learning continuous signed distance functions for shape
  representation.
\newblock In {\em Proceedings of the IEEE Conference on Computer Vision and
  Pattern Recognition}, pages 165--174, 2019.

\bibitem{qi2017pointnet}
Charles~R Qi, Hao Su, Kaichun Mo, and Leonidas~J Guibas.
\newblock Pointnet: Deep learning on point sets for 3d classification and
  segmentation.
\newblock In {\em Proceedings of the IEEE conference on computer vision and
  pattern recognition}, pages 652--660, 2017.

\bibitem{ravi2020accelerating}
Nikhila Ravi, Jeremy Reizenstein, David Novotny, Taylor Gordon, Wan-Yen Lo,
  Justin Johnson, and Georgia Gkioxari.
\newblock Accelerating 3d deep learning with pytorch3d.
\newblock {\em arXiv preprint arXiv:2007.08501}, 2020.

\bibitem{remelli2020meshsdf}
Edoardo Remelli, Artem Lukoianov, Stephan~R. Richter, Benoit Guillard, Timur~M.
  Bagautdinov, P. Baqu{\'e}, and P. Fua.
\newblock Meshsdf: Differentiable iso-surface extraction.
\newblock {\em ArXiv}, abs/2006.03997, 2020.

\bibitem{runz2020frodo}
Martin Runz, Kejie Li, Meng Tang, Lingni Ma, Chen Kong, Tanner Schmidt, Ian
  Reid, Lourdes Agapito, Julian Straub, Steven Lovegrove, et~al.
\newblock Frodo: From detections to 3d objects.
\newblock In {\em Proceedings of the IEEE/CVF Conference on Computer Vision and
  Pattern Recognition}, pages 14720--14729, 2020.

\bibitem{schwarz2020graf}
Katja Schwarz, Yiyi Liao, Michael Niemeyer, and Andreas Geiger.
\newblock Graf: Generative radiance fields for 3d-aware image synthesis.
\newblock {\em arXiv preprint arXiv:2007.02442}, 2020.

\bibitem{sitzmann2019deepvoxels}
Vincent Sitzmann, Justus Thies, Felix Heide, Matthias Nie{\ss}ner, Gordon
  Wetzstein, and Michael Zollhofer.
\newblock Deepvoxels: Learning persistent 3d feature embeddings.
\newblock In {\em Proceedings of the IEEE/CVF Conference on Computer Vision and
  Pattern Recognition}, pages 2437--2446, 2019.

\bibitem{sitzmann2020scene}
Vincent Sitzmann, Michael Zollhöfer, and Gordon Wetzstein.
\newblock Scene representation networks: Continuous 3d-structure-aware neural
  scene representations, 2020.

\bibitem{pix3d}
Xingyuan Sun, Jiajun Wu, Xiuming Zhang, Zhoutong Zhang, Chengkai Zhang, Tianfan
  Xue, Joshua~B Tenenbaum, and William~T Freeman.
\newblock Pix3d: Dataset and methods for single-image 3d shape modeling.
\newblock In {\em IEEE Conference on Computer Vision and Pattern Recognition
  (CVPR)}, 2018.

\bibitem{tatarchenko2015single}
Maxim Tatarchenko, Alexey Dosovitskiy, and Thomas Brox.
\newblock Single-view to multi-view: Reconstructing unseen views with a
  convolutional network.
\newblock {\em CoRR abs/1511.06702}, 1(2):2, 2015.

\bibitem{tatarchenko2019single}
Maxim Tatarchenko, Stephan~R Richter, Ren{\'e} Ranftl, Zhuwen Li, Vladlen
  Koltun, and Thomas Brox.
\newblock What do single-view 3d reconstruction networks learn?
\newblock In {\em Proceedings of the IEEE Conference on Computer Vision and
  Pattern Recognition}, pages 3405--3414, 2019.

\bibitem{trevithick2020grf}
Alex Trevithick and Bo Yang.
\newblock Grf: Learning a general radiance field for 3d scene representation
  and rendering.
\newblock {\em arXiv preprint arXiv:2010.04595}, 2020.

\bibitem{wang2018pixel2mesh}
Nanyang Wang, Yinda Zhang, Zhuwen Li, Yanwei Fu, Wei Liu, and Yu-Gang Jiang.
\newblock Pixel2mesh: Generating 3d mesh models from single rgb images.
\newblock In {\em Proceedings of the European Conference on Computer Vision
  (ECCV)}, pages 52--67, 2018.

\bibitem{wang2004image}
Zhou Wang, Alan~C Bovik, Hamid~R Sheikh, and Eero~P Simoncelli.
\newblock Image quality assessment: from error visibility to structural
  similarity.
\newblock {\em IEEE transactions on image processing}, 13(4):600--612, 2004.

\bibitem{wang2021nerf}
Zirui Wang, Shangzhe Wu, Weidi Xie, Min Chen, and Victor~Adrian Prisacariu.
\newblock Nerf $--$: Neural radiance fields without known camera parameters.
\newblock {\em arXiv preprint arXiv:2102.07064}, 2021.

\bibitem{3dgan}
Jiajun Wu, Chengkai Zhang, Tianfan Xue, William~T Freeman, and Joshua~B
  Tenenbaum.
\newblock Learning a probabilistic latent space of object shapes via 3d
  generative-adversarial modeling.
\newblock In {\em Advances in Neural Information Processing Systems}, pages
  82--90, 2016.

\bibitem{wu2019detectron2}
Yuxin Wu, Alexander Kirillov, Francisco Massa, Wan-Yen Lo, and Ross Girshick.
\newblock Detectron2.
\newblock \url{https://github.com/facebookresearch/detectron2}, 2019.

\bibitem{wu20153d}
Zhirong Wu, Shuran Song, Aditya Khosla, Fisher Yu, Linguang Zhang, Xiaoou Tang,
  and Jianxiong Xiao.
\newblock 3d shapenets: A deep representation for volumetric shapes.
\newblock In {\em Proceedings of the IEEE conference on computer vision and
  pattern recognition}, pages 1912--1920, 2015.

\bibitem{xu2019disn}
Qiangeng Xu, Weiyue Wang, Duygu Ceylan, Radomir Mech, and Ulrich Neumann.
\newblock Disn: Deep implicit surface network for high-quality single-view 3d
  reconstruction.
\newblock In {\em Advances in Neural Information Processing Systems}, pages
  492--502, 2019.

\bibitem{yen2020inerf}
Lin Yen-Chen, Pete Florence, Jonathan~T Barron, Alberto Rodriguez, Phillip
  Isola, and Tsung-Yi Lin.
\newblock inerf: Inverting neural radiance fields for pose estimation.
\newblock {\em arXiv preprint arXiv:2012.05877}, 2020.

\bibitem{yu2020pixelnerf}
Alex Yu, Vickie Ye, Matthew Tancik, and Angjoo Kanazawa.
\newblock pixelnerf: Neural radiance fields from one or few images.
\newblock {\em arXiv preprint arXiv:2012.02190}, 2020.

\end{thebibliography}
}

\end{document}